\documentclass[epj]{webofc}
\usepackage[utf8]{inputenc}
\usepackage[varg]{txfonts}   
\usepackage{booktabs}
\usepackage{xcolor}
\definecolor{darkred}{rgb}{0.4,0.0,0.0}
\definecolor{darkgreen}{rgb}{0.0,0.4,0.0}
\definecolor{darkblue}{rgb}{0.0,0.0,0.4}
\usepackage[bookmarks,linktocpage,colorlinks,
    linkcolor = darkred,
    urlcolor  = darkblue,
    citecolor = darkgreen]{hyperref}
\usepackage{subfigure}
\usepackage{bbold}
\wocname{EPJ Web of Conferences}
\woctitle{Lattice2017}
\begin{document}
%
\selectlanguage{english}
\title{%
Computation of parton distributions from the quasi-PDF \\ approach at the physical point
}
\author{%
\firstname{Constantia} \lastname{Alexandrou}\inst{1,2} \and
\firstname{Simone} \lastname{Bacchio}\inst{1,3}  \and
\firstname{Krzysztof} \lastname{Cichy}\inst{4,5} \and
\firstname{Martha} \lastname{Constantinou}\inst{6} \and
\firstname{Kyriakos} \lastname{Hadjiyiannakou}\inst{2} \and
\firstname{Karl} \lastname{Jansen}\inst{7} \and
\firstname{Giannis} \lastname{Koutsou}\inst{2} \and
\firstname{Aurora} \lastname{Scapellato}\inst{1,3}\fnsep\thanks{Speaker, \email{scapellato.aurora@ucy.ac.cy}} \and 
\firstname{Fernanda} \lastname{Steffens}\inst{7}
}
\institute{%
Department of Physics, University of Cyprus, POB 20537, 1678 Nicosia, Cyprus
\and
The Cyprus Institute, 20 Kavafi Str., Nicosia 2121, Cyprus
\and
University of Wuppertal, Gaußstr. 20, 42119 Wuppertal, Germany
\and
Goethe-Universit\"{a}t Frankfurt am Main, Institut f\"{u}r Theoretische Physik, Max-von-Laue-Strasse 1, 60438 Frankfurt am Main, Germany
\and
Faculty of Physics, Adam Mickiewicz University, Umultowska 85, 61-614 Pozna\'{n}, Poland
\and
Department of Physics, Temple University, Philadelphia, PA 19122 - 1801, USA
\and
John von Neumann Institute for Computing (NIC), DESY, Platanenallee 6, 15738 Zeuthen, Germany
}
\abstract{%
We show the first results for parton distribution functions  within the proton at the physical pion mass, employing the method of quasi-distributions. In particular, we present the matrix elements for the iso-vector combination of the unpolarized, helicity and transversity quasi-distributions,  obtained with $N_f=2$ twisted mass clover-improved fermions and a proton boosted with momentum $\vert \vec{p}\vert=0.83\mbox{ GeV}$. The momentum smearing technique has been applied to improve the overlap with the proton boosted state. Moreover, we present the renormalized helicity matrix elements  in the RI$'$ scheme, following the non-perturbative renormalization prescription recently developed by our group.
}
\begin{flushright}
DESY 17-160 
\end{flushright}
\maketitle
\section{Introduction}\label{intro}
Parton distribution functions (PDFs) provide an important insight into the inner dynamics of hadrons, by giving information on the momentum distribution of their constituents (quarks, antiquarks and gluons), and represent also essential inputs for making predictions in high energy experiments.
Our knowledge on PDFs relies so far on phenomenological analysis of deep inelastic scattering data (see for instance Refs.~\citep{Owens:2012bv, Alekhin:2013nda, deFlorian:2009vb, Ball:2017nwa}) and suffers from a number of shortcomings.
Therefore, finding an alternative approach for computing PDFs would be a major step. Indeed, from phenomenological fits, a strong dependence of the result has been observed on the form of the parametrization and on the selected data; moreover, the scarcity of experimental data in certain kinematic regions leads to very poorly constrained distributions especially at large and small values of the Bjorken variable $x$. On the other hand, an \textit{ab initio} direct determination of parton distributions from QCD has been impossible until the pioneering work of Ji to compute the  quasi-PDFs within the lattice QCD framework~\cite{Ji:2013dva}.
The standard lattice QCD methods could not be used   since PDFs are defined in terms of non-local light cone correlations in the Minkowskian space-time, usually in the rest frame of the nucleon. In practice, if we simulate QCD in  Euclidean space, we cannot reproduce light cone distances, because these would require a zero lattice spacing. This issue is overcome by the so-called quasi-PDF approach. The idea is to compute correlation functions in which the quark-antiquark fields are separated by space-like distances and connected by a Wilson line that guarantees the gauge invariance of the relevant matrix element. The validity of this method relies on the nucleon momentum boost, which has to be as large as possible in order to re-establish a safe connection to the physical PDFs. The latter is pursued by a perturbative matching procedure.

Obtaining a good signal when using highly boosted states is not a trivial issue in lattice QCD, because the two- and three-point functions, that are needed for computing the PDFs, decay exponentially with the energy of the proton. To address this problem, a momentum dependent smearing has been proposed very recently~\cite{Bali:2016lva} and a significant improvement of the quality of the signal for PDFs has been seen when compared to the one coming from only the standard Gaussian smearing~\cite{Alexandrou:2016jqi}. Many aspects of the quasi-PDF approach have been investigated so far and among these we mention the matching procedure up to one-loop perturbation theory~\cite{Xiong:2013bka,Chen:2016fxx,Wang:2017qyg,Stewart:2017tvs} and the target mass corrections (TMCs)~\cite{Alexandrou:2015rja,Chen:2016utp}. Very recently,  the renormalization pattern of non-local matrix elements for quasi-PDFs has been addressed and presented in Refs.~\cite{Constantinou:2017sej,Alexandrou:2017huk,Green:2017xeu}.

At present, Ji's approach has been employed for two different $N_f=2+1+1$ lattice QCD ensembles at a non- physical pion mass. The results for non-singlet isovector operators for unpolarized, helicity and transversity distributions look very promising, as can be seen from the bare distributions that move towards the physical PDFs~\cite{Alexandrou:2016jqi,Lin:2014zya,Alexandrou:2015rja,Chen:2016utp} when increasing the momentum of the proton. There are still many aspects that need to be investigated, because the distributions extracted so far from the lattice suffer from several systematic effects, deriving for instance from a finite lattice spacing, non-physical quark masses and lattice artifacts in the renormalization functions. The systematic effect we want to address in this work is the dependence of quasi-PDFs on the pion mass.

\section{Lattice calculation}
According to Ji's approach, quasi-PDFs are computed from
\begin{equation}
\tilde{q}(x,\Lambda,P_i)=\int _{-\infty}^{\infty}\, \frac{dz}{4\pi}e^{-izxP_i}\langle P\vert \bar{\psi}(0)\Gamma_i W(0,z)\psi(z)\vert P\rangle\quad ,
\label{eq:def}
\end{equation}
where $\Lambda$ ($\sim 1/a$, in a lattice calculation) is the UV cutoff and $\vert P\rangle$ is the proton state with momentum $P=(P_0,P_1,P_2,P_3)$. We take only one spatial component of the momentum different from zero (denoted here by $P_i$, where $i$ can be $1,2,3$) and the Wilson line $W(0,z)$ aligned to the direction of motion, that we call $i$-direction. The quantity $xP_i$ can be interpreted as the quark momentum in the direction of the boost. Choosing specific $\Gamma_i$ matrices in the insertion operator, we can study three types of PDFs, namely unpolarized, helicity and transversity PDFs. These give information about the momentum distribution of partons with a given spin projection along the direction of the motion. 

\noindent
In this work we use:
\begin{itemize}
\item $\Gamma_i=\gamma_i$ for the unpolarized case ($\tilde{q}=\tilde{q}_{\downarrow}+\tilde{q}_{\uparrow}$)
\item $\Gamma_i=\gamma_5\gamma_i$ for the helicity case ($\Delta \tilde{q}=\tilde{q}_{\downarrow}-\tilde{q}_{\uparrow}$)
\item $\Gamma_i=\sigma_{ij}$ for the transversity case ($\delta \tilde{q}=\tilde{q}_{\perp}-\tilde{q}_{\top}$)
\end{itemize}
The simulations are performed with a $48^3\times 96$ ensemble at approximately physical pion mass ($m_\pi=131$ MeV), produced by ETMC~\cite{Abdel-Rehim:2015pwa}, with $N_f=2$ flavours of maximally twisted mass clover-improved fermions, bare coupling of $\beta=2.10$, corresponding to a lattice spacing $a\simeq 0.093$~fm and twisted mass parameter $a\mu=0.0009$. We apply $N_G=50$ iterations of Gaussian smearing on the quark fields with $\alpha=4$ and $N_{\rm APE}=50$ iterations of APE smearing, with $\alpha_{APE}=0.5$. The results presented in this work have been produced at the boost momentum $\vert \vec{p}\vert=6\pi/L$, corresponding to $0.83$ GeV. To improve the quality of the signal, we tune the momentum smearing parameter $\zeta$, computing the nucleon two point-functions using $50$ configurations. The momentum smearing function on the quark fields, $\psi(x)$, reads,
\begin{equation}
S_{mom}\, \psi(x)=\frac{1}{6\alpha}\left( \psi(x)+\alpha\sum_j U_j(x) e^{i\zeta\hat{e}_j}\psi(x+\hat{e}_j)\right) \quad .
\end{equation}
The optimal value for $\zeta$ was found to be $0.6$. We kept $0.6$ also for the matrix elements of Eq.\eqref{eq:def}.

To increase statistics and minimize lattice artifacts
 we compute quasi-PDFs with the proton  boosted along $x$-, $y$- or $z$- axis and also in the positive or negative orientation, for a total of six possible combinations. However, changing either direction or orientation of the momentum requires a new inversion of the Dirac matrix, because the gauge links are multiplied by a different phase factor. To calculate the three-point functions, we employ the sequential method for the all-to-all propagator with the fixed sink approach, setting the source-sink separation at $t_s=12 a\simeq 1.11 \mbox{ fm}$. Thus, we compute $6$ up/down forward propagators and $24$ up/down backward propagators (considering that, for transversity quasi-PDFs, two choices for $\sigma_{ij}$ can contribute if the proton is moving along the $i$-direction).
For our computation with momentum $6\pi/L$, we analyze $60$ configurations with $16$ source positions each, for a total of $5760$ measurements for the unpolarized and helicity and $11520$ for transversity distributions.

\section{Results}
\begin{figure}[thb]
  \centering
  \includegraphics[width=6.65cm,clip]{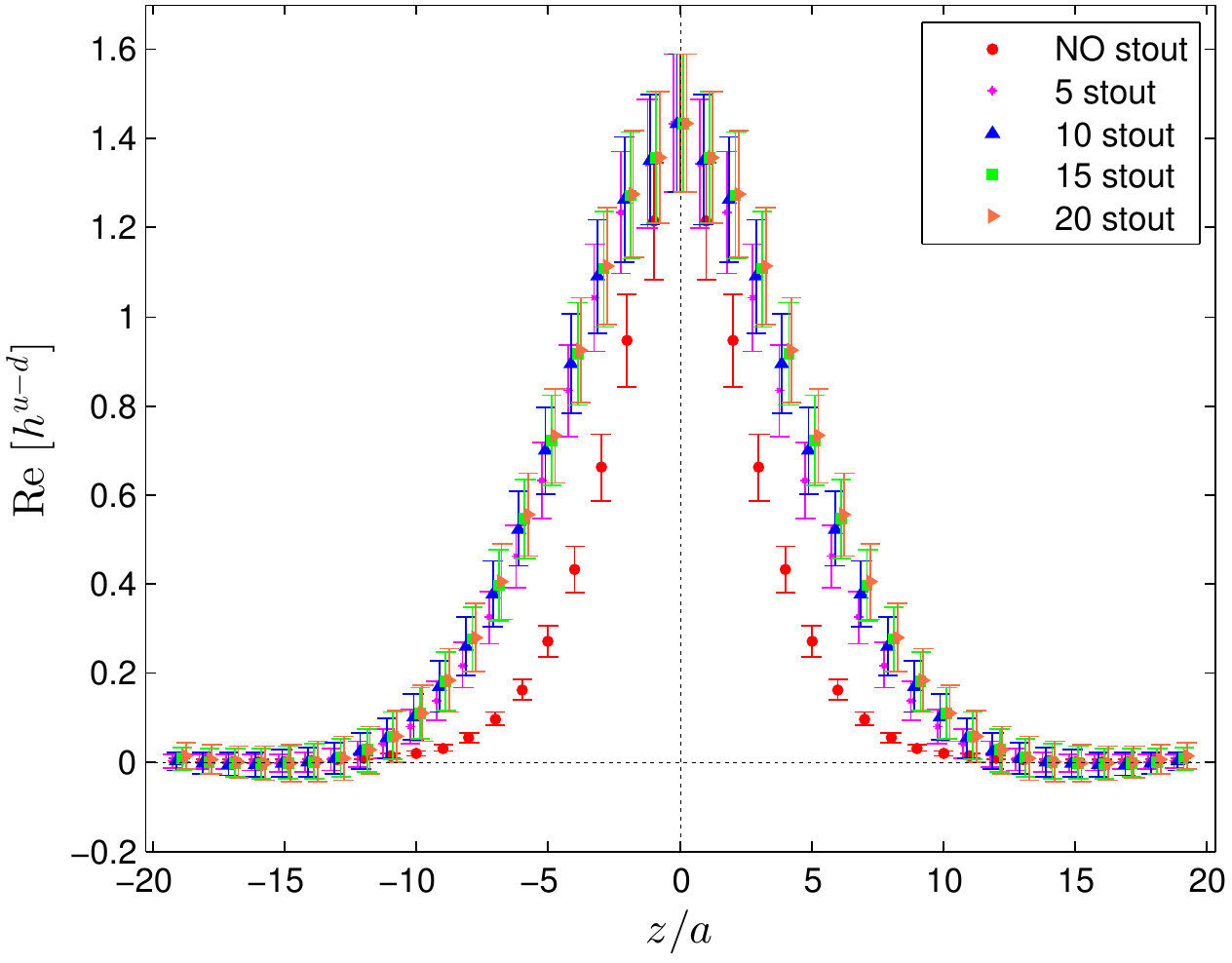}
  \qquad
   \includegraphics[width=6.65cm,clip]{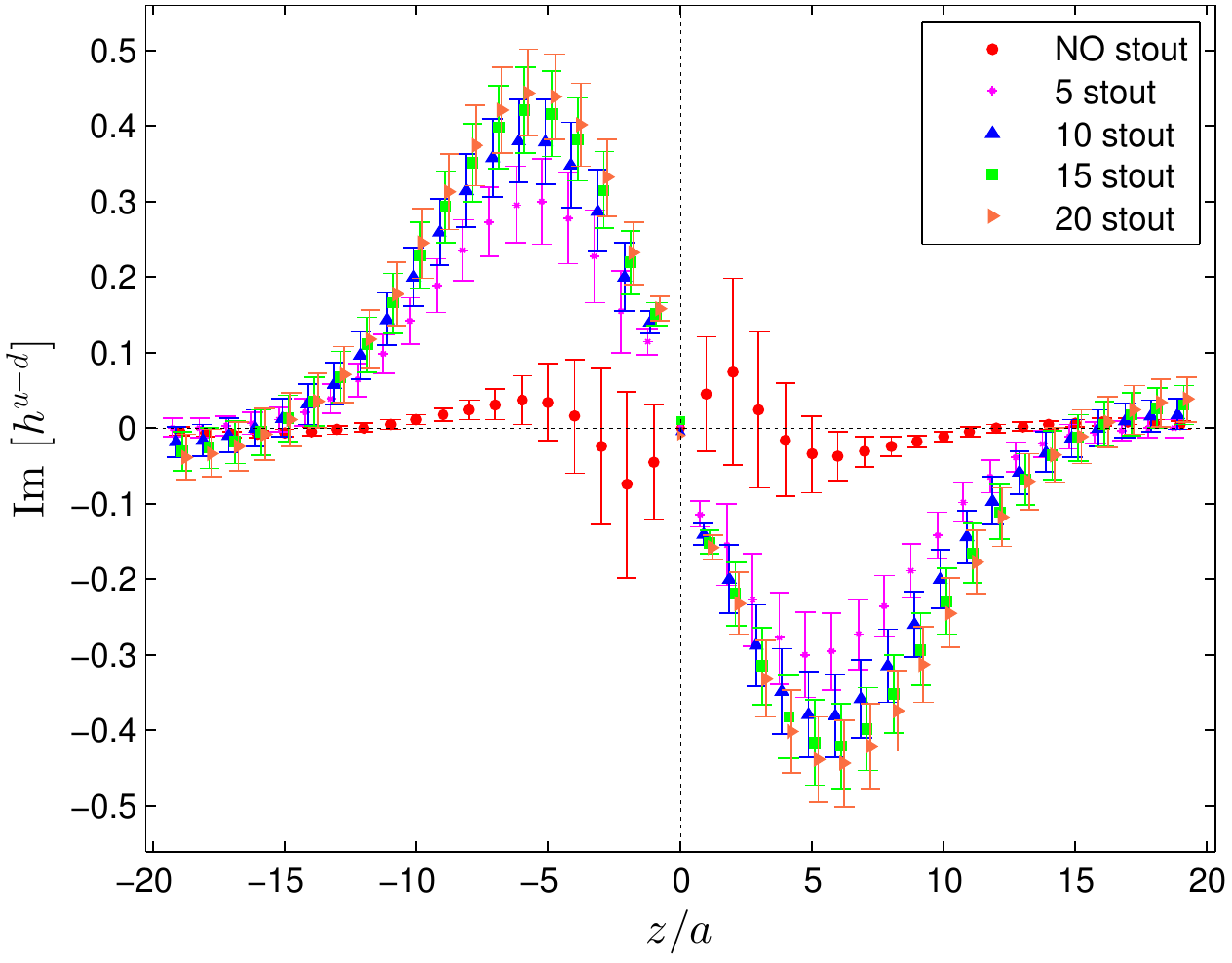}
  \caption{Real (left panel) and imaginary part (right panel) of the vector form factors for different stout steps and $\vert \vec{p}\vert=6\pi/L$.}
  \label{fig:unpol}
\end{figure}

\begin{figure}[thb]
  \centering
  \includegraphics[width=6.65cm,clip]{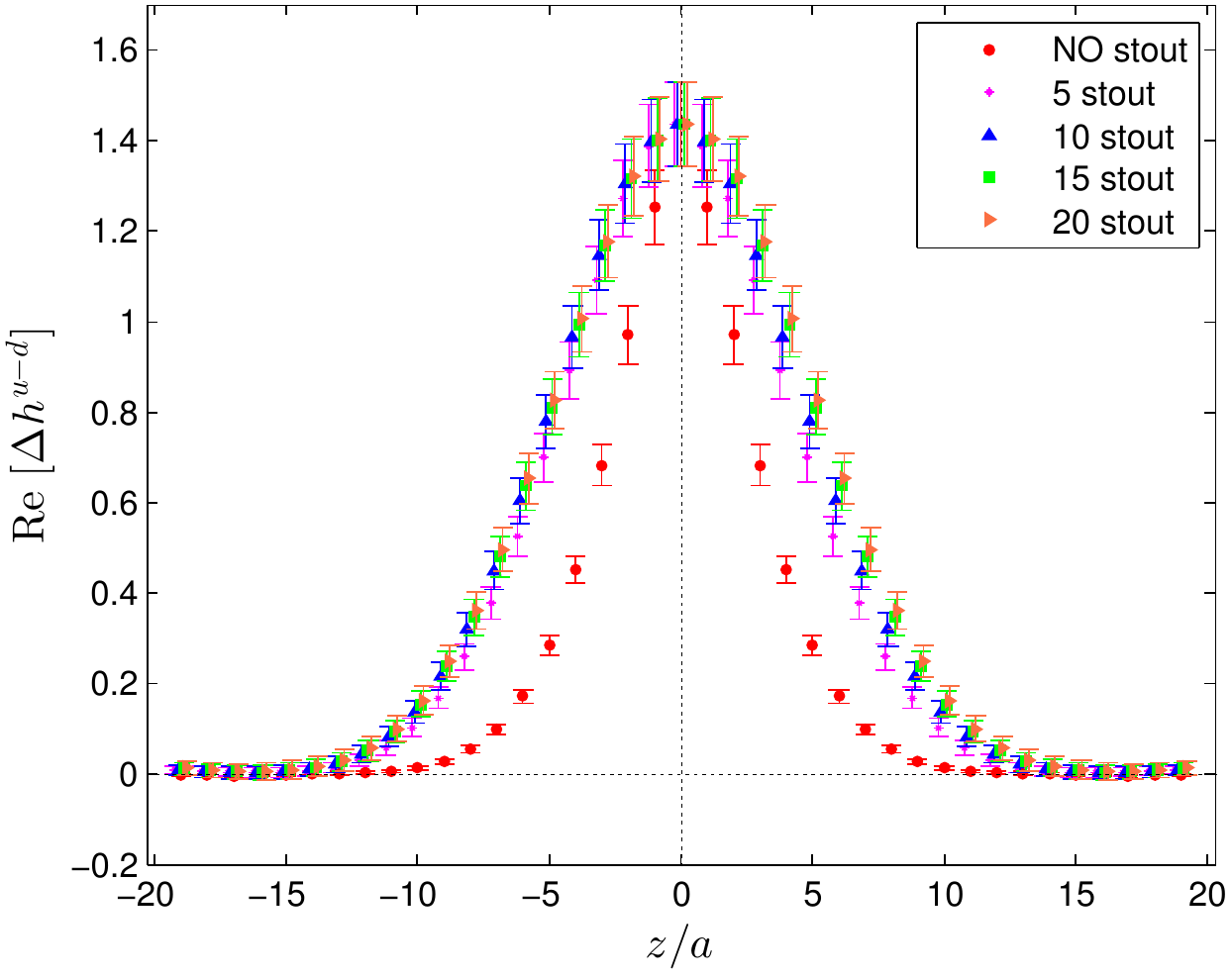}
  \qquad
   \includegraphics[width=6.65cm,clip]{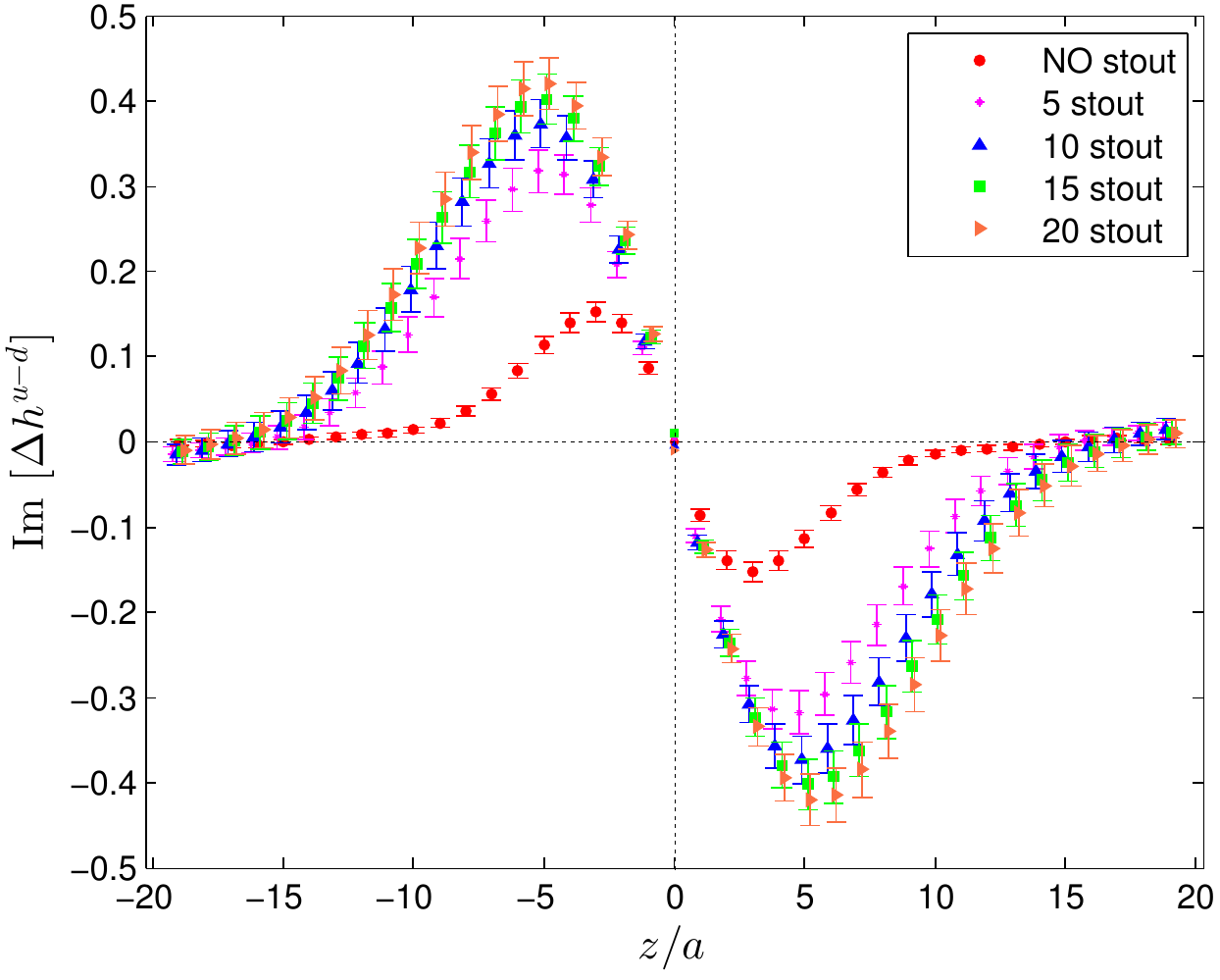}
  \caption{Real (left panel) and imaginary part (right panel) of the axial-vector form factors for different stout steps and $\vert \vec{p}\vert=6\pi/L$.}
  \label{fig:hel}
\end{figure}

\begin{figure}[thb]
  \centering
  \includegraphics[width=6.65cm,clip]{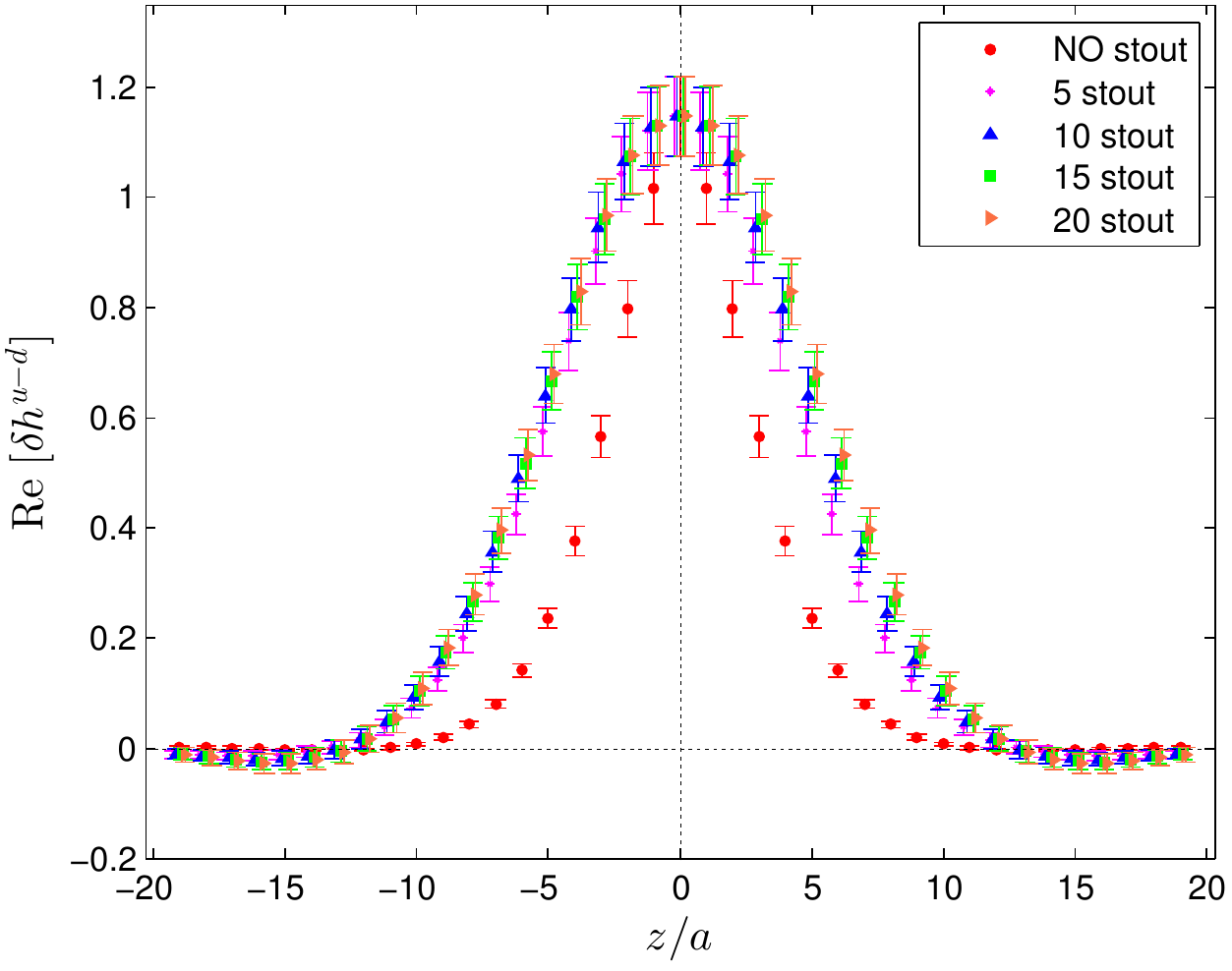}
  \qquad
   \includegraphics[width=6.65cm,clip]{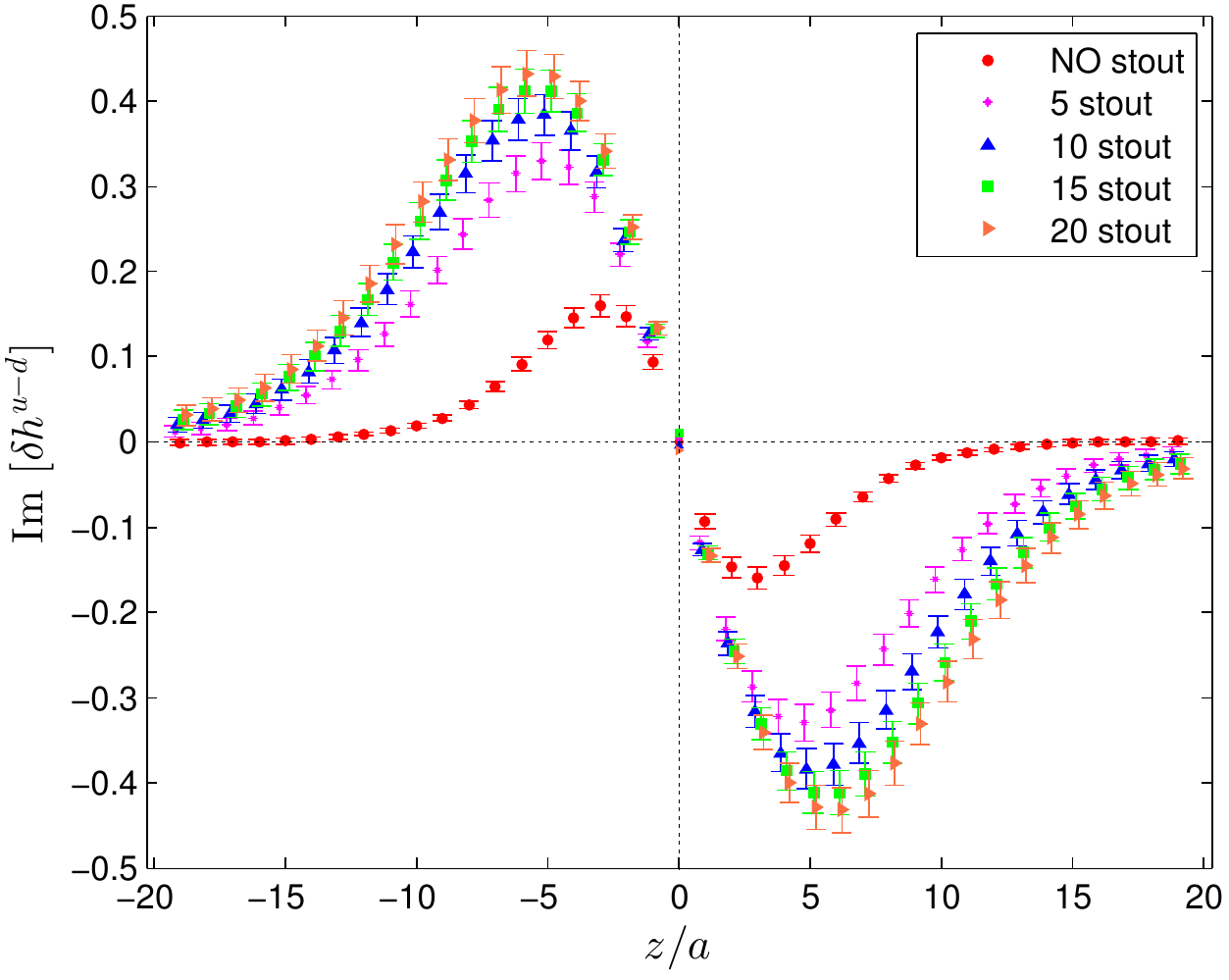}
  \caption{Real (left panel) and imaginary part (right panel) of the tensor form factors for different stout steps and $\vert \vec{p}\vert=6\pi/L$.}
  \label{fig:tra}
\end{figure}

 We show the unrenormalized matrix elements for the unpolarized, helicity and transversity cases in Figs.~\ref{fig:unpol}, \ref{fig:hel} and \ref{fig:tra}, relevant for the isovector combination $u$-$d$. This choice is motivated by the possibility of avoiding disconnected quark loops, which are present for instance in operators of the type $O=\overline{\psi}\Gamma \mathbb{1}\psi$, where $\mathbb{1}$ is the identity matrix in the flavour space. In order to estimate the effect of the renormalization and bring the renormalization factors closer to their tree level value, we apply multiple steps of stout smearing to the gauge links in the operator. The matrix elements have been saved every $5$ sweeps of stout smearing, from $0$ up to $20$ steps.

\begin{figure}[thb]
  \centering
  \includegraphics[width=7cm,clip]{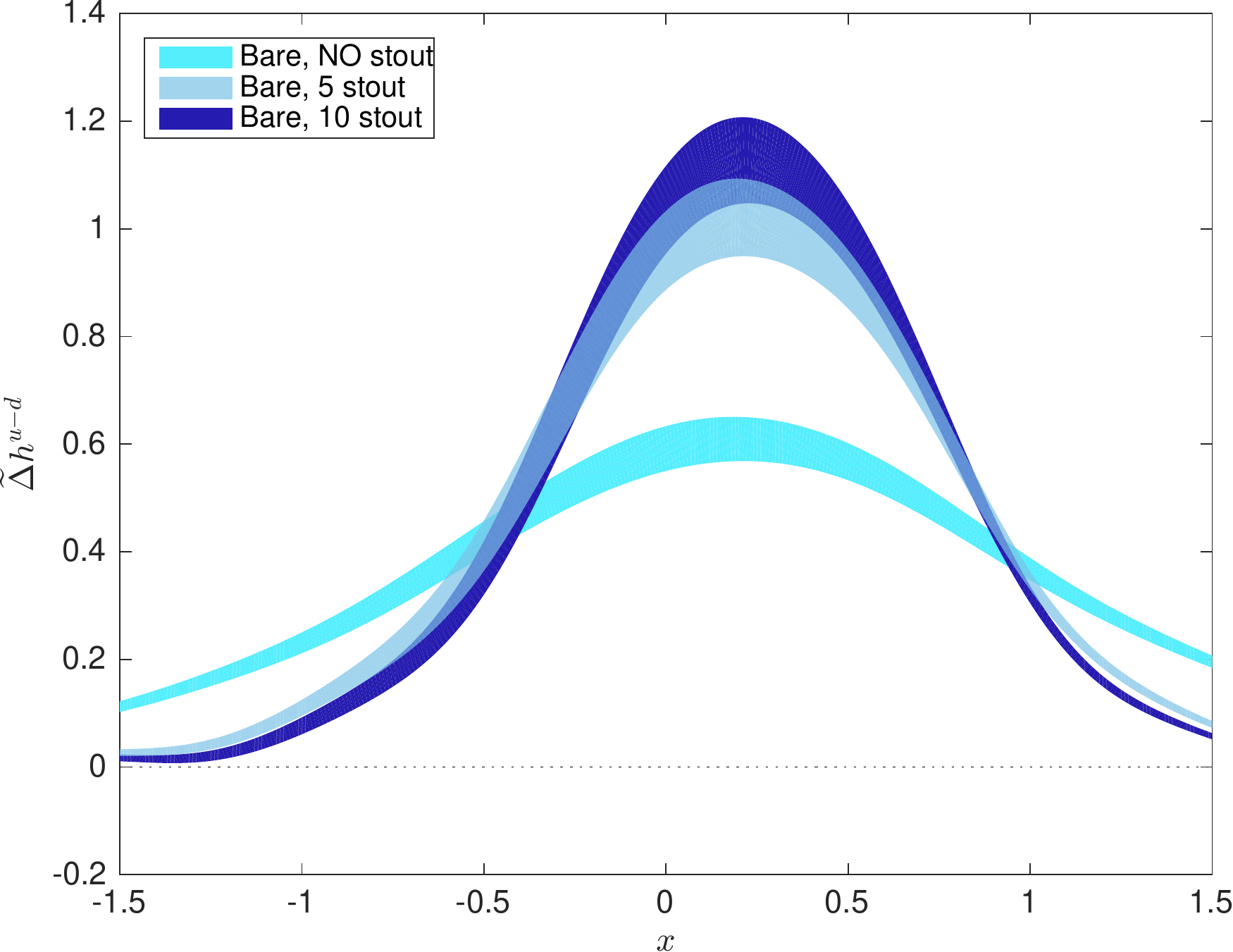}
 \caption{Comparison of bare helicity quasi-distributions for $0$, $5$ and $10$ iterations of stout smearing.}
  \label{fig:trasf_bare}
\end{figure}

It can be observed that applying stout smearing to the Wilson line clearly affects the value of the bare matrix elements. In general, the impact of smearing on the imaginary part seems considerably stronger than on the real part. When comparing the different numbers of smearing steps, the change from zero to five steps appears to be more significant than from five to ten steps, suggesting a saturation of the smearing effect.
As pointed out in Ref.~\cite{Alexandrou:2015rja},  smearing is used to explore the effect of renormalization on quasi-PDFs. The enhancement of the imaginary part with the stout smearing leads to a strong $x$-asymmetry in the bare quasi-distribution, as illustrated in Fig.~\ref{fig:trasf_bare} for the helicity case. This reveals the expected quark-antiquark asymmetry in the distribution, since antiquarks can be interpreted as quarks in the negative $x$ region, according to the crossing relation $\overline{\Delta}q(-x)=\Delta \bar{q}(x)$.

\begin{figure}[thb]
  \centering
  \includegraphics[width=7.35cm]{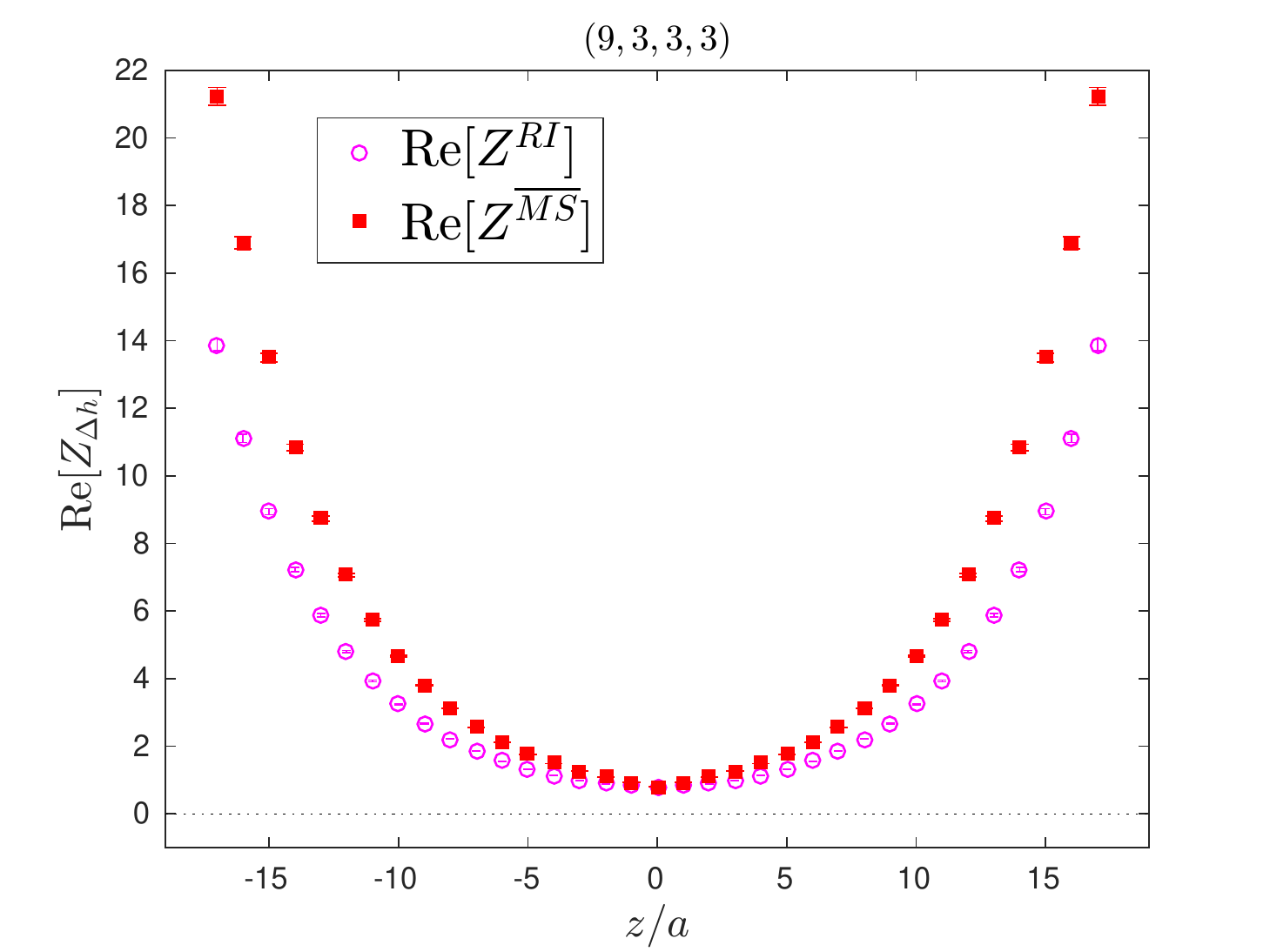}
  \quad
  \includegraphics[width=6.35cm]{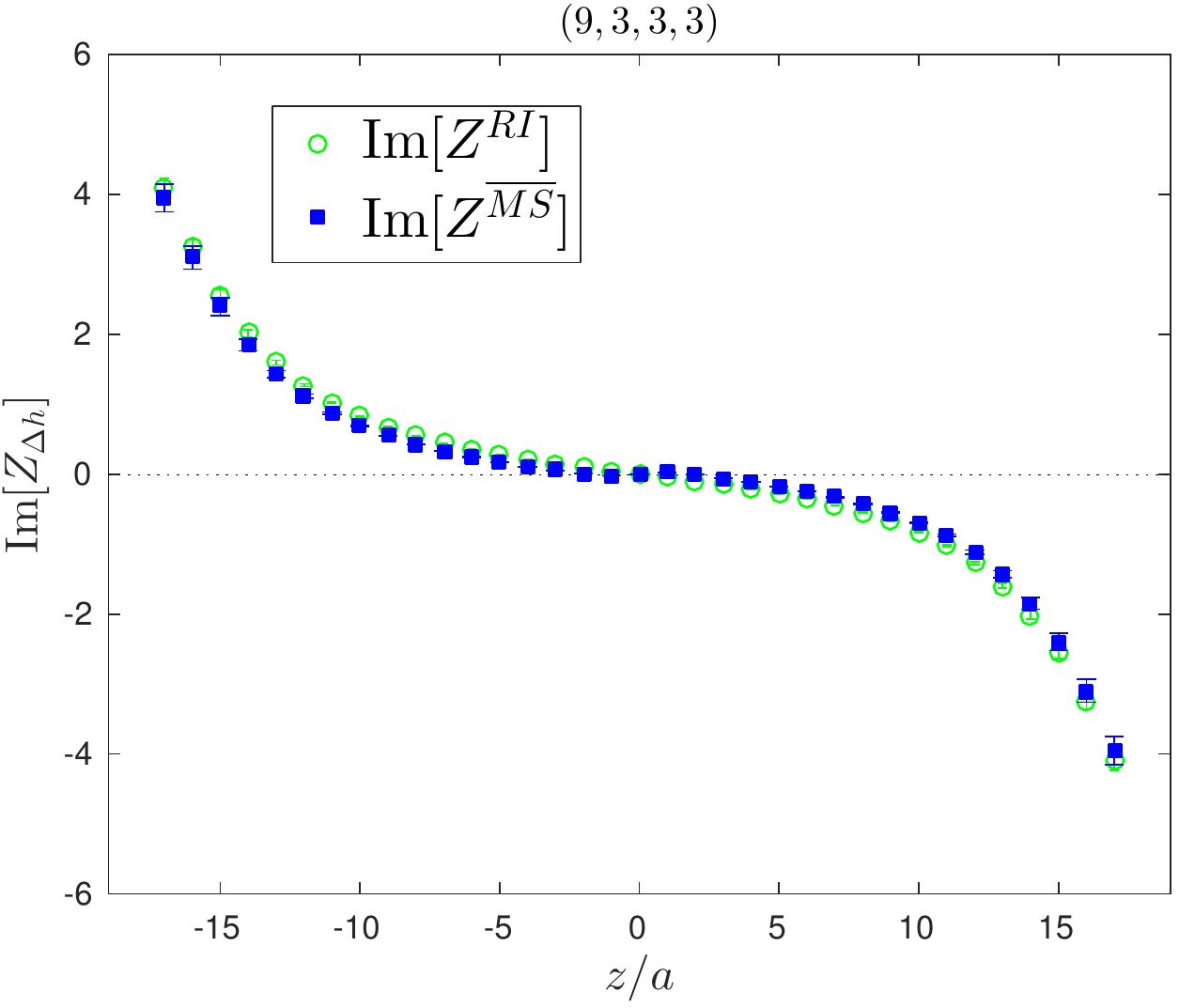}
  \caption{Real (left panel) and imaginary parts (right panel) of the renormalization factors as a function of the length of the Wilson line, $z$, for $20$ steps of stout smearing. Open (filled) symbols correspond to the RI$'$ ($\overline{\mbox{MS}}$) scale estimates.}
  \label{fig:zfactor}
\end{figure}

A valuable cross-check of the lattice QCD results can be done by examining the matrix elements at $z = 0$, which can be identified at $Q^2 = 0$ with the local vector, axial and tensor charges. Using the Z-factors for local operators in the $\overline{\mbox{MS}}$ scheme at $\mu=2$ GeV listed in~\cite{Alexandrou:2015sea}, we find for $\vert \vec{p}\vert=6\pi/L$: $Z_{V}h^{u-d}(0)= 1.09(12)$, $Z_{A} \Delta h^{u-d}(0) = 1.136(74)$, $Z_{T} \delta h^{u-d}( 0) = 0.981(62)$. These values are compatible within errors with the local vector charge $g_V^{u-d}$, which has to be $1$ for the proton and the axial and tensor charges $g_A^{u-d}$, $g_T^{u-d}$ computed on this ensemble~\cite{Alexandrou:2017hac,Alexandrou:2017qyt}.
Following the guidelines of the renormalization programme presented in~\cite{Alexandrou:2017huk} and~\cite{Alexandrou:2017qpu}, we apply the non-perturbative renormalization prescription on the matrix elements corresponding to the helicity PDF. We will address in the future the renormalization of the unpolarized and transversity distributions. As shown for the first time in Ref.~\cite{Constantinou:2017sej}, for the helicity there is no mixing with other operators when the index of the Dirac matrix in the insertion is in the same direction as the Wilson line. The renormalization programme is based on a RI$'$ scheme~\cite{Martinelli:1994ty} and on the following renormalization conditions for the Wilson-line operator $Z_O$ and quark field $Z_q$:
\begin{equation}
Z_q^{-1}Z_O(z)\frac{1}{12}\mbox{Tr}\left[ \mathcal{V}(p,z)(\mathcal{V}^{Born}(p,z))^{-1}\right]\Big{\vert}_{p^2 =\bar{\mu}^2_{0}}=1 \quad ,
\label{eq:zq}
\end{equation} 
\begin{equation}
Z_q=\frac{1}{12}\mbox{Tr}\left[ (S(p)^{-1}) S^{Born}(p)\right] \Big{\vert}_{p^2 =\bar{\mu}^2_{0}}  \quad .
\label{eq:zo}
\end{equation}
In Eqs.\eqref{eq:zq} and~\eqref{eq:zo}, $\mathcal{V}(p,z)$ and $S(p)$ stand for the amputated vertex function of the  axial-vector operator  and the quark propagator, respectively, while $\mathcal{V}^{Born}(p,z)$ and $S^{Born}(p)$ correspond to their tree-level values, with $\mathcal{V}^{Born}(p,z)=i\gamma_i\gamma_5 e^{ipz}$ for the helicity operator. For the evaluation of the vertex functions we adopt the momentum source technique~\cite{Gockeler:1998ye}, which allows us to extract the renormalization functions for any operators, at each value $z$ of the Wilson line length, computing only one quark propagator momentum dependent~\cite{Alexandrou:2010me}.
The momentum used in the source determines the RI$'$  renormalization scale $\bar{\mu}_0$, having the form $\bar{\mu_0}=\frac{2\pi}{L}\left( \frac{n_0}{2}+\frac{1}{4},n_1,n_2,n_3\right)$, where $L$ is the spatial extent of the lattice. To renormalize our results, we use a ``diagonal'' RI$'$ scale for the momenta (i.e. all spatial components $P_i$ are different from zero) since it is expected to have smaller lattice artifacts~\citep{Alexandrou:2015sea,Alexandrou:2017huk}. Keeping in mind these two requirements, we set the scale at $a\bar{\mu}_0=\frac{2\pi}{48}\left( \frac{9}{2}+\frac{1}{4},3,3,3\right)$. The temporal component has been chosen in order to make $(ap)^2$ large enough so that the non-perturbative effects are not very strong and $\hat{P}=\frac{\sum _\rho \bar{{\mu}_0}_\rho^4}{\left( \sum _\rho \bar{{\mu}_0}_\rho^2\right) ^2}$ sufficiently small to avoid enhanced cut-off effects (see Refs.~\cite{Alexandrou:2010me,Constantinou:2010gr} and~\cite{Alexandrou:2017huk} for a specific study on the quasi-PDFs).
In Fig.~\ref{fig:zfactor} we show the extracted values for the real and imaginary part of helicity Z-factor, $Z_{\Delta h}$, for $20$ steps of stout smearing. In each plot, we report the results in the RI$'$ scheme and the Z-factors converted to $\overline{\mbox{MS}}$ scheme at $\bar{\mu}=2$ GeV. The conversion factor has been computed in one-loop continuum perturbation theory in~\citep{Constantinou:2017sej} and the evolution to the $\overline{\mbox{MS}}$ scale $\bar{\mu}=2$ GeV is performed using the intermediate Renormalization Group Invariant scheme (RGI), employed in~\cite{Alexandrou:2015sea} and~\cite{Constantinou:2014fka}.
As can be observed, the errors on the Z-factors are very tiny and this is an advantage of the momentum source method, which offers a high statistical accuracy with a small number of measurements. We also note that the values of the Z-factors increase with increasing $z$ as a consequence of the power divergences of long-link operators. Moreover, the imaginary part is considerably smaller than the real part (both in RI$'$ and $\overline{\mbox{MS}}$ schemes) and this is expected from perturbation theory, where the renormalization functions are real in dimensional regularization and $\overline{\mbox{MS}}$ scheme to all orders.

\begin{figure}
  \centering
  \includegraphics[width=6.6cm,clip]{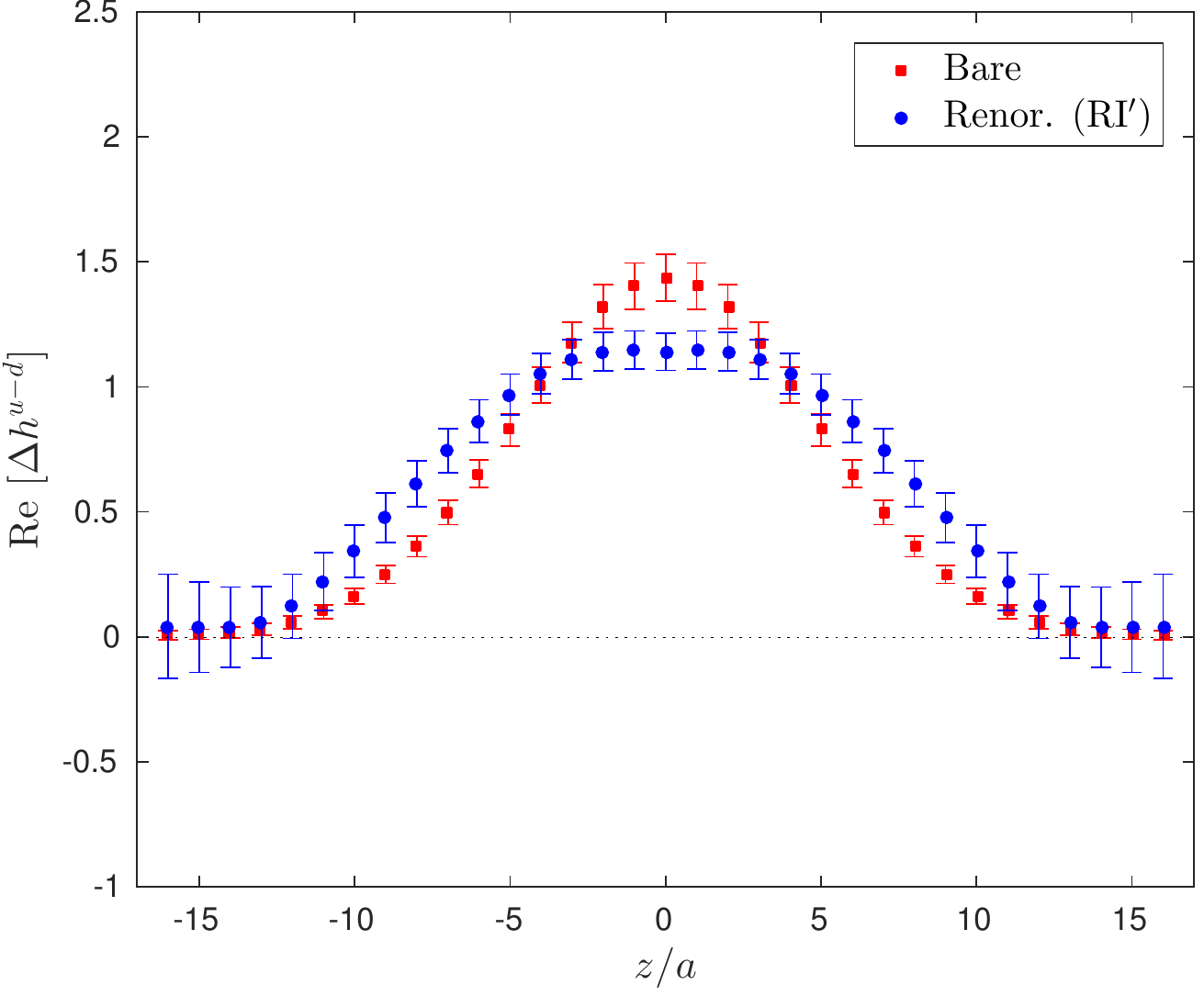}
  \qquad
   \includegraphics[width=6.6cm,clip]{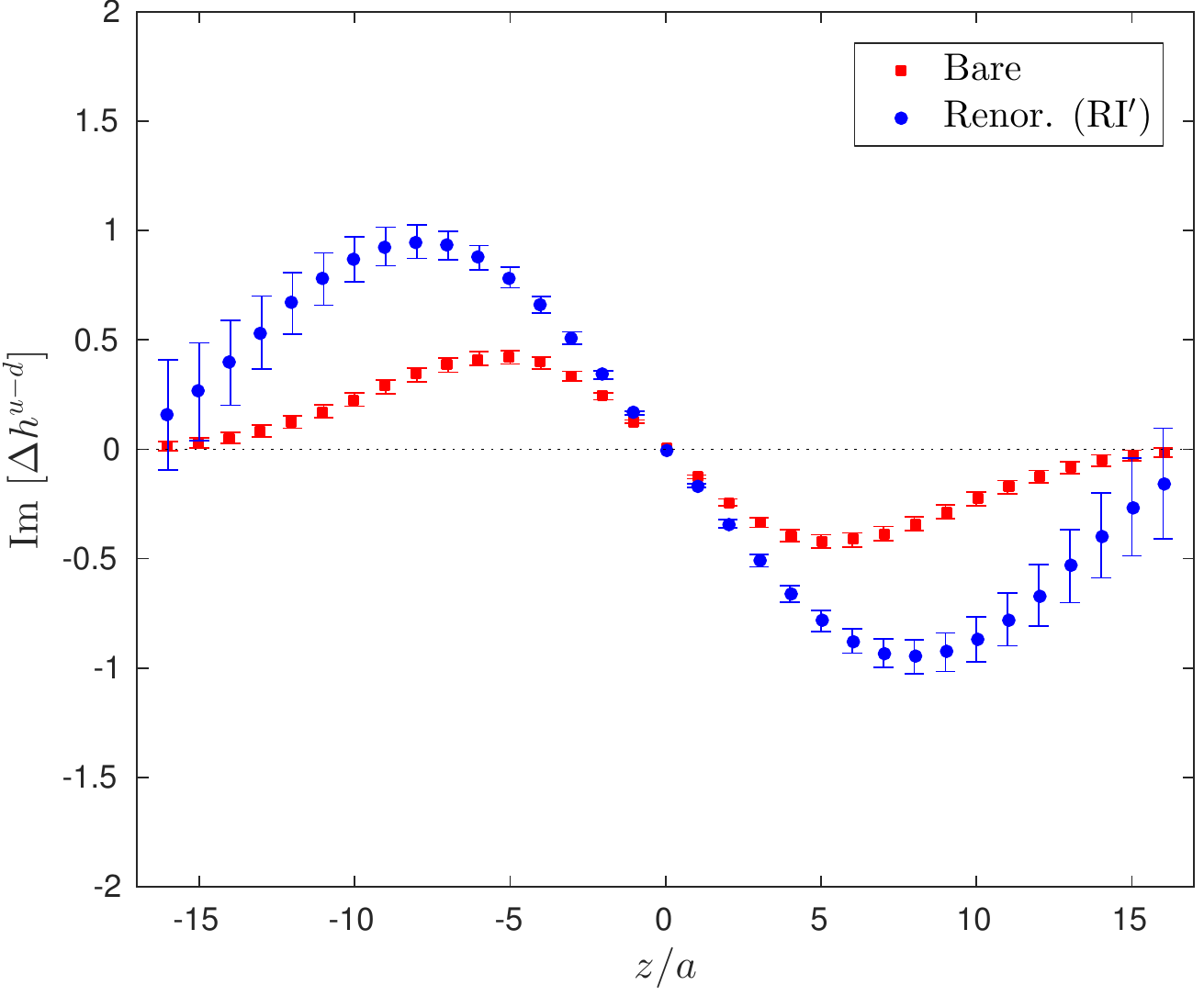}
  \caption{Renormalized matrix elements for the helicity quasi-PDFs in the RI$'$ scheme at scale $a\bar{\mu}_0=\left( \frac{9}{2}+\frac{1}{4},3,3,3\right) $ for $20$ stout smearing steps.}
  \label{fig:hel_renorm}
\end{figure}

Having the Z-factors, we renormalize the matrix elements for the helicity distribution multiplying each matrix element $\Delta h(P,z)$ by its renormalization function. In Fig.~\ref{fig:hel_renorm} we show the real and imaginary parts of the renormalized matrix elements in RI$'$ scheme, as a result of the complex multiplication $\Delta h(P,z)^{RI'}=\Delta h(P,z) ^{bare}\times  Z_{\Delta h}^{RI'}(P,z)$. From Fig.~\ref{fig:hel_renorm} one can see that Re[$\Delta h^{RI'}$] is compatible with zero from $z/a>12$, but with increased statistical uncertainties. Moreover, after renormalization, the imaginary part is amplified especially for intermediate values of $z$ and this is mostly due to the fact that Im[$\Delta h^{RI'}$] receives significant contributions from the term Re[$Z_{\Delta h}^{RI'}$] Im[$\Delta h^{bare}$].
\vspace*{-0.5cm}
\vspace*{0.35cm}
\section{Conclusions and perspectives}
In this work, we report first results on the PDFs using an ensemble of twisted mass fermions simulated at the physical pion mass. As a starting point we set the momentum of the proton to $6\pi/L$, which corresponds to $0.83$ GeV in physical units. This is still a very low value for the momentum, where we know there cannot be a quantitative agreement with phenomenological expectations and neither a safe matching to the physical PDFs. Nevertheless, the calculation for such a low momentum gives us an indication about the computational effort needed to extract the PDFs from a physical point ensemble, since this work represents the first attempt in this direction. It is indeed interesting to make a comparison with the previous study on the PDFs with momentum smearing~\cite{Alexandrou:2016jqi}, where we used a $32^3\times 64$ ensemble from ETMC, with $N_f=2+1+1$ at $M_\pi\approx 375$ MeV~\cite{Baron:2010bv}. In that work, $150$ measurements were enough to have a precision of $10\%$ with momentum $6\pi/L$ ($1.41$ GeV in physical units), while at the physical point we need around $38$ times more measurements for the same accuracy and with a lower momentum. However, we are currently computing quasi-PDFs using larger boosts, which is a highly non-trivial task due to  the increased noise-to-signal ratio.
The bare matrix elements for the unpolarized, helicity and transversity distributions need to be renormalized. The renormalization functions for the helicity distribution, $Z_{\Delta h}$, are computed in a RI$'$ and $\overline{\mbox{MS}}$ schemes, and allow us to extract the renormalized matrix elements for the helicity case. In future work we will address the renormalization of the unpolarized distribution, taking into account the mixing with the scalar operator in the twisted basis and the renormalization of the transversity distribution, which proceeds in complete analogy with the helicity one. We should also keep in mind that, also for the unpolarized PDF, there is the possibility to avoid the operator mixing if in the insertion one uses an appropriate set of gamma matrices instead of the standard choice of a $\gamma$ matrix in the direction of the Wilson line~\cite{Constantinou:2017sej}. The practicality of this approach needs to be explored. Moreover, the explicit computation of lattice artifacts in the renormalization functions and the calculation of the conversion factor RI$'$-$\overline{\mbox{MS}}$ up to two-loops are also important ingredient towards a more reliable estimate of PDFs from the lattice.

\section*{Acknowledgements}
We would like to thank all members of ETMC for their constant and pleasant collaboration. We also thank Haralambos Panagopoulos for his contribution to the renormalization of quasi-PDFs. K.C. was supported in part by the Deutsche Forschungsgemeinschaft (DFG), project nr. CI 236/1-1. MC acknowledges financial support by the U.S. Department of Energy, Office of Nuclear Physics, within the framework of the TMD Topical Collaboration, as well as, by the National Science Foundation under Grant No. PHY-1714407. This research used computational resources provided by the Titan supercomputer at the Oak Ridge Leadership Computing Facility (OLCF).
The quark propagators   have been computed using the DD$\alpha$AMG method optimized for twisted mass fermions~\cite{Alexandrou:2016izb}.
This work has  received funding from the European Union's Horizon 2020 research and innovation programme under the Marie Sk\l{}odowska-Curie grant agreement No 642069 (HPC-LEAP).

\clearpage
\bibliography{lattice2017}

\end{document}